\documentclass{iau} 
\usepackage{graphicx}

\title[IAUS 346.~~The dark side of sgHMXB] 
{The dark side of \\ supergiant High-Mass X-ray Binaries}

\author[S. Chaty, F. Fortin \& F. Garc\'ia \& F. Fogantini]   
{Sylvain Chaty$^1$,
  Francis Fortin$^1$,
  Federico Garc\'ia$^1$
  \and Federico Fogantini$^2$}

\affiliation{$^1$AIM, CEA, CNRS, Universit\'e Paris-Saclay, Universit\'e Paris Diderot, Sorbonne Paris Cit\'e, F-91191 Gif-sur-Yvette, France \\ email: {\tt chaty@cea.fr} \\[\affilskip]
  $^2$Instituto Argentino de Radioastronom\'ia (CCT-La Plata, CONICET, CICPBA), C.C. No. 5, 1894 Villa Elisa, and Facultad de Ciencias Astron\'omicas y Geofisicas, Universidad Nacional de La Plata, Paseo del Bosque s/n, 1900 La Plata, Argentina}

\pubyear{2018}
\volume{346}  
\jname{High-mass X-ray binaries: illuminating the passage from massive binaries 
to merging compact objects}
\editors{L. Oskinova, ed.}

\def\Lsol{\mbox{ }L_{\odot}}
\def\Mdot{\frac{dM}{dt}}
\def\Msol{\mbox{ }M_{\odot}}
\def\Mstar{\mbox{ }M_{\star}}
\def\Rsol{\mbox{ }R_{\odot}}
\def\Rstar{\mbox{ }R_{\star}}
\def\ergs{\mbox{ erg\,s}^{-1}}
\def\nh{N_{\rm H}}
\def\cmmoinsdeux{\mbox{ cm}^{-2}}
\def\Porb{P_{orb}}
\def\Pspin{P_{spin}}
\def\adeg{^{\circ}}

\begin{document}

\maketitle

\begin{abstract}
  High Mass X-ray Binaries (HMXB) have been revealed by a wealth of multi-wavelength observations, from X-ray to optical and infrared domain. After describing the 3 different kinds of HMXB, we focus on 3 HMXB hosting supergiant stars: IGR~J16320-4751, IGR~J16465-4507 and IGR~J16318-4848, respectively called ``The Good'', ``The Bad'' and ``The Ugly''.
  We review in these proceedings what the observations of these sources have brought to light concerning our knowledge of HMXB, and what part still remains in the dark side. Many questions are still pending, related to accretion processes, stellar wind properties in these massive and active stars, and the overall evolution due to transfer of mass and angular momentum between the companion star and the compact object. Future observations should be able to answer these questions, which constitute the dark side of HMXB.
\end{abstract}

\firstsection 
\section{Introduction}

Intensive programs, including imaging, photometry, low and high resolution spectroscopy, stellar spectra modeling, spectral energy distribution (SED) fitting, timing and interferometry, have shown that properties of HMXB are mainly dictated by the nature of their massive host stars. Imaging and photometry allow us to identify various types of HMXB; low and high resolution spectroscopy, combined to stellar spectra modeling, lead us to derive accurate parameters of the companion star (interstellar absorption, metallicity, rotation, gravity, etc); SED fitting gives us information on intrinsic absorption and characteristics of circumstellar enveloppe; mid-infrared imaging allows us to explore the impact of these active stars on their environment; timing brings us orbital and spin periods; and finally interferometry opens the way to directly imaging the dust cocoon surrounding HMXB. The {\it INTEGRAL} satellite has triggered the revival of HMXB studies, extending the population of supergiant HMXB --from only 5 in 1986 to 35 today--, revealing previously unknown highly obscured and transient HMXB (so-called supergiant Fast X-ray Transients, SFXT). The first detections of gravitational waves has confirmed the interest of studying compact binaries hosting massive stars, the obscured HMXB being the precursors of common enveloppe systems.

\section{Introduction on HMXB systems}

``High-Mass X-ray Binaries'' (HMXB) are composed of a compact object (neutron star --NS-- or black hole --BH--) orbiting a luminous and massive early OB spectral type companion star ($\geq 10 \Msol$). We can distinguish 3 different types of HMXB, according to the process of accretion, that we order here by decreasing number of systems of each class: (i) Be X-ray binaries (BeHMXB), (ii) supergiant X-ray binaries (sgHMXB), and (iii) Roche Lobe Overflow systems (RLO) (see more extensive discussion in \cite[Tauris et al. 2017]{tauris:2017}). \\

\begin{figure}
  \begin{center}
    \includegraphics[width=5.75in]{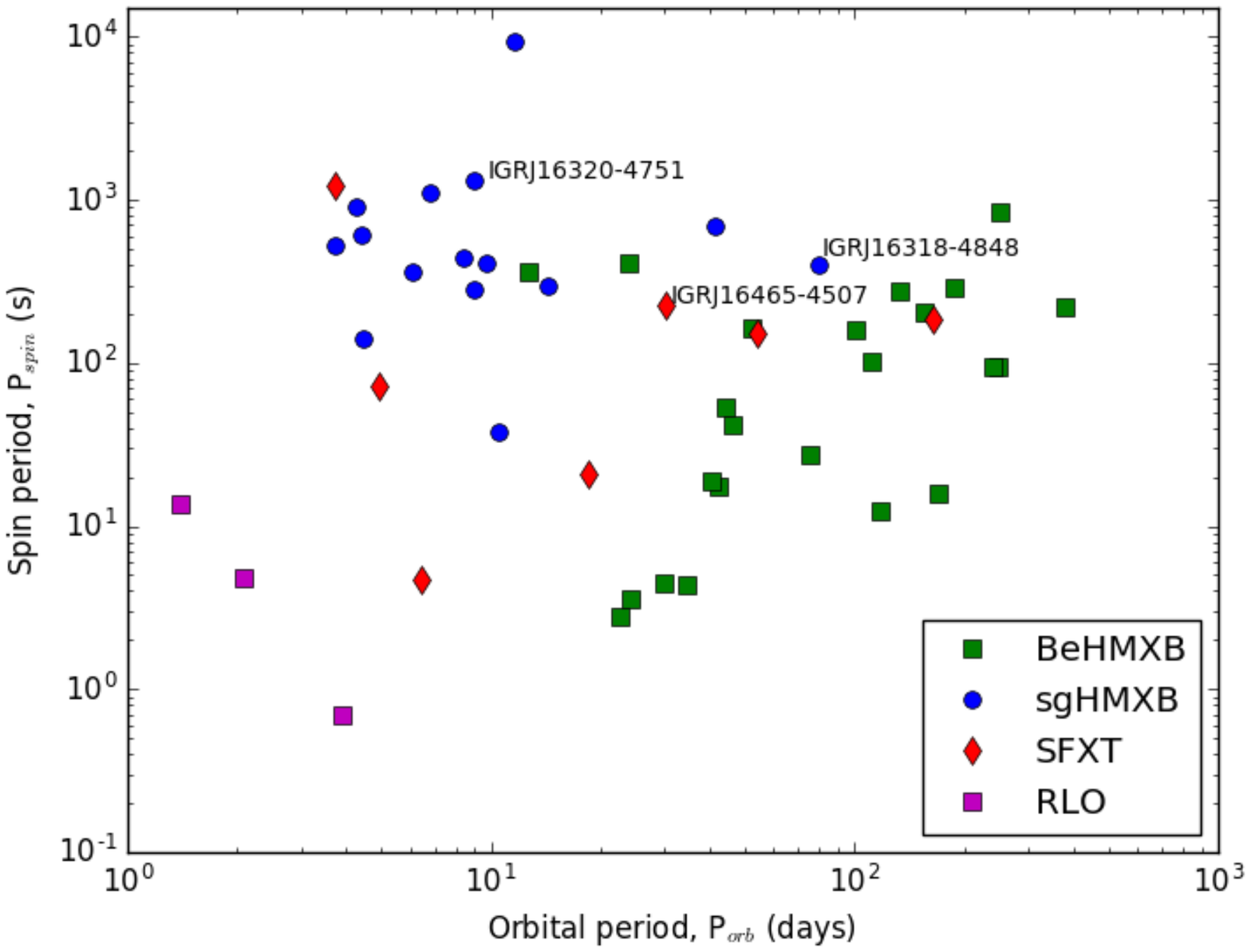}
    \caption{Corbet diagram showing the different populations of HMXB. We indicated the positions of IGR~J16320-4751 and IGR~J16465-4507. Concerning IGR~J16318-4848, since we only know the orbital period (80d), we indicate the position of the source, assuming an average spin period for sgHMXB.}
    \label{fig:corbet}
  \end{center}
\end{figure}

{\underline{\it (i) Be X-ray binaries (BeHMXB)}}

BeHMXB host a main sequence donor star of spectral type B0-B2e~III/IV/V, a rapid rotator surrounded by a circumstellar (so-called ``decretion'') disc of gas, as seen by the presence of a prominent H$\alpha$ emission line. This disc is created by a low velocity/high density stellar wind of $\sim 10^{-7} \Msol / yr$. Transient and bright (``Type I'') X-ray outbursts periodically occur, each time the compact object (usually a NS on a wide and eccentric orbit) approaches periastron and accretes matter from the decretion disc (see \cite[Charles \& Coe 2006, Tauris \& van den Heuvel 2006]{charles:2006,tauris:2006} and references therein). These systems exhibit a correlation between the spin and orbital period, as shown by their location from the lower left to the upper right of the Corbet diagram (initial diagram shown in \cite[Corbet 1986]{corbet:1986}, updated diagram in Fig.~\ref{fig:corbet}), due to efficient transfer of angular momentum at each periastron passage: rapidly spinning NS correspond to short orbital period systems, and slowly spinning NS to long orbit systems. Apart from MWC\,656 hosting a BH (\cite[Casares et al. 2014]{casares:2014}), most BeHMXB seem to host a NS. \\

{\underline{\it (ii) Supergiant X-ray binaries (sgHMXB)}}

sgHMXB host a supergiant star of spectral type O8-B1~I/II, characterized by an intense, slow and dense, radiatively steady and highly supersonic stellar wind, radially outflowing from the equator. There are $\sim 16$ so-called ``classical'' sgHMXB, most of them being close systems, with a compact object orbiting on a short and circular orbit, directly accreting from the stellar wind, through e.g. Bondy-Hoyle-Littleton process. Such wind-fed systems exhibit a luminous and persistent X-ray emission ($L_X = 10^{36-38} \ergs$), with superimposed large variations on short timescales, and a cut-off (10-30 keV) power-law X-ray spectrum. Located in the upper left part of the Corbet diagram, with small orbital period $P_\mathrm{orb} \sim 3-10$\,days and long spin periods $P_\mathrm{spin} \sim 100-10000$\,s (see Fig.~\ref{fig:corbet}), they do not show any correlation, due to absence of net transfer of angular momentum. Nearly half of sgHMXB ($\sim 8$) exhibit a substantial intrinsic and local extinction $\nh \geq 10^{23} \cmmoinsdeux$, with a compact object deeply embedded in the dense stellar wind (such as the highly obscured IGR~J16318-4848; \cite[Chaty \& Rahoui 2012]{chaty:2012a}). Likely in transition to Roche Lobe Overflow, these systems are characterized by slow winds causing a deep spiral-in of the compact object, and leading to Common Envelope Phase. Detection of long pulsations imply that they host young NS with $B \sim 10^{11-12}$\,G. There exists also the possibility, for sgHMXB such as Cyg~X-1, to accrete both through Roche lobe overflow and stellar wind accretion.

A significant subclass of sgHMXB is constituted of 17 (+5 candidate) Supergiant Fast X-ray Transients (SFXT, see \cite[Negueruela et al. 2006]{negueruela:2006a}). These systems, characterized by a compact object orbiting with P$_{orb} \sim 3.3 - 100$\,days on a circular or excentric orbit, and by $100-1000$\,s spin periods, span on a vast location in the Corbet diagram, mostly inbetween BeHMXB and sgHMXB (see Fig.~\ref{fig:corbet}). They exhibit short and intense X-ray outbursts, an unusual characteristic among HMXB, rising in tens of minutes up to a peak luminosity $L_X \sim 10^{35-37} \ergs$, lasting for a few hours, and alternating with long ($\sim 70$\,days) quiescence at $L_X \sim 10^{32-34} \ergs$, with an impressive variability factor $\frac{L_{max}}{L_{min}}$ going up to $10^2-10^5$. Various processes have been invoked to explain these flares, such as wind inhomogeneities, magneto/centrifugal accretion barrier, transitory accretion disc, etc (see the reviews \cite[Chaty 2013]{chaty:2013} and \cite[Walter 2015]{walter:2015}, and references therein). \\

{\underline{\it (iii) (Beginning Atmospheric) Roche Lobe Overflow systems (RLO)}}

RLO host a massive star filling its Roche lobe, where accreted matter flows via inner Lagrangian point to form an accretion disc (similarly to LMXB). These systems are also called beginning atmospheric Roche lobe overflow. They constitute the classical «bright» HMXB (such as Cen X-3, SMC X-1 and LMC X-4), with accretion of matter occuring through the formation of an accretion disc, leading to a high X-ray luminosity ($L_X \sim 10^{38} \ergs$) during outbursts. There are only a few sources, located in the lower left of the Corbet diagram (Fig.~\ref{fig:corbet}), characterized by short orbital and spin periods. \\

A number of 114 HMXB are reported in \cite[Liu et al. (2006)]{liu:2006}, and 117 in \cite[Bird et al. (2016)]{bird:2016}. By cross-correlating both catalogues, we find that they share 79 sources in common. Among these common sources, 6 sources identified as HMXB in \cite[Liu et al. (2006)]{liu:2006} are now assigned to other types by \cite[Bird et al. (2016)]{bird:2016}. We therefore find that the total number of HMXB currently known in our Galaxy amounts to 152 \cite[(Fortin et al. 2018a)]{fortin:2018a}, see also this volume \cite[(Fortin et al. (2018b)]{fortin:2018b}. HMXB thus represent $\sim 40\%$ of the total number of high energy binary systems (i.e. adding all known LMXB and HMXB reported in \cite[Liu et al. 2007 and Bird et al. 2016]{liu:2007,bird:2016}). Among HMXB, there are 63 BeHMXB (51 firmly identified and 12 candidates), 33 sgHMXB (30 firmly identified and 3 candidates) and 56 HMXB of unidentified nature. sgHMXB can be further divised in 16 ``classical'' sources and 17 SFXT. Thus, HMXB can be divided respectively in 41\% of BeHMXB, 22\% of sgHMXB and 37\% of unidentified HMXB. \\

Contrary to LMXB located towards the Galactic center, HMXB are concentrated in the Galactic plane, towards tangential directions of Galactic arms, rich in star forming regions and stellar formation complexes (\cite[SFC, Coleiro \& Chaty 2013]{coleiro:2013a} and references therein). The distribution of HMXB in our Galaxy is thus a good indicator of starburst activity, and their collective X-ray luminosity has been used to compute the star-formation rate of the host galaxy \cite[(Grimm et al. 2003)]{grimm:2003}. \cite[Coleiro \& Chaty (2013)]{coleiro:2013a} have correlated the position of HMXB (including BeHMXB and sgHMXB) with the position of SFC, and showed that HMXB are clustered within 0.3\,kpc of the closest SFC, with an inter-cluster distance of 1.7\,kpc, thus showing that HMXB remain close to their birthplace.
\cite[Coleiro \& Chaty (2013)]{coleiro:2013a} have also shown that the HMXB distribution was offset by $\sim 10^7$\,years with respect to the spiral arms, corresponding to the delay between star birth and HMXB formation. By taking into account the galactic arm rotation, they managed to derive parameters such as age, migration distance and kick for some BeHMXB and sgHMXB.

\section{``The Good'': IGR~J16320-4751}

IGR~J16320-4751 is a highly obscured sgHMXB hosting an O8~I star, with column density $\nh = 10^{23} \cmmoinsdeux$. We named it ``The Good'' source, because it is located well inside the sgHMXB domain in the Corbet diagram (Fig.~\ref{fig:corbet}), with $\Pspin = 1300$\,s and $\Porb = 9$\,days. We studied this source aiming at constraining both its geometrical and physical properties, following the evolution of the NS orbiting the supergiant star. We give here a summary of the study described in detail in \cite[Garc\'ia et al. (2018a)]{garcia:2018a}, and also in this volume \cite[Garc\'ia et al. (2018b)]{garcia:2018b}.

To study this source, we retrieved i.) the {\it Swift}/BAT folded hard X-ray lightcurve from 2004 to 2017, and ii.) the {\it XMM}-Newton/PN lightcurves (soft 0.5-6.0 keV and hard 6.0-12.0 keV bands), from 11 observations spanning from 2003 to 2008. The source exhibits a high variability and flaring activity on several timescales. For instance, in one of the XMM observations, we found two flares with an increase of a factor $\sim 10$ in only 300\,s. Using the {\it Swift}/BAT lightcurve, we refined the orbital period to $8.99 \pm 0.01$~days.

The {\it XMM}-Newton PN spectra of the source exhibit a highly absorbed continuum, with an Fe absorption edge at $\sim 7$\,keV, along with Fe\,K$\alpha$, Fe\,K$\beta$ and Fe\,XXV lines. We performed a spectral fitting, with thermally comptonized COMPTT model, adding three narrow Gaussian functions with 2 TBABS absorption components. We then extracted the line characteristics, the total continuum flux and the intrinsic $\nh$.

We first found a clear correlation between Fe\,K$\alpha$ line and the total continuum flux, showing that the fluorescence emission emanates from a small region close to the accreting pulsar. We then found another correlation between the Fe\,K$\alpha$ line flux and the intrinsic $\nh$, suggesting that the fluorescent matter is linked to the absorbing matter. Both results taken together show that absorbing matter is located within a small dense region surrounding the NS, likely due to accreted stellar wind.

\begin{figure}
  \begin{center}
    \includegraphics[width=3.75in,angle=-90]{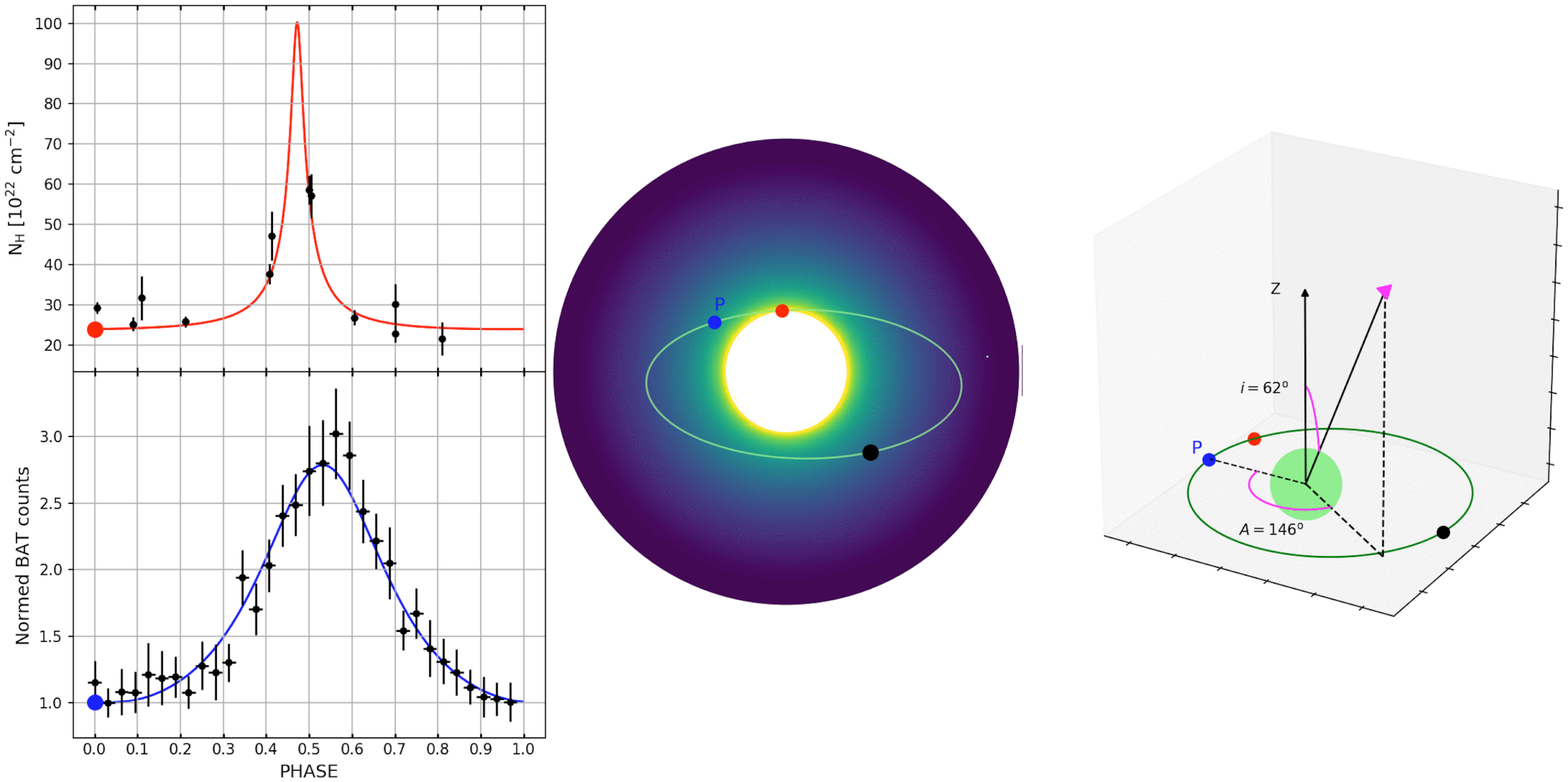}
    \caption{Top left: In black, data points showing the variation of $\nh$ along orbital phase, in red: model fit. Bottom left: in black, data points showing the variation of normalised {\it Swift}/BAT counts along orbital phase, in blue: model fit. Middle: supergiant star surrounded by NS as seen by observer, Periastron point (P) indicated, the red point is the furthest point on the line-of-sight. Right: our simple model, with NS (in black) orbiting the supergiant star (in green), periastron (P) indicated, the red point is the furthest point on the line-of-sight, the pink arrow indicates the direction of the observer. i is the inclination angle, and A the azimuth.}
    \label{fig:igrj16320}
  \end{center}
\end{figure}

We show in Fig.~\ref{fig:igrj16320} the orbital evolution of the intrinsic absorption $\nh$ derived from fitting {\it XMM}-Newton spectra (top left), and the orbital evolution of the folded hard X-ray {\it Swift}/BAT lightcurve (bottom left). In order to fit $\nh$ and {\it Swift}/BAT lightcurves, we built a toy model of a NS orbiting a supergiant star, and accreting from its stellar wind (\cite[Castor et al. 1975]{castor:1975} model, red lines): $\Mstar = 25 \Msol$, $\Rstar = 20 \Rsol$, $\beta = 0.85$, $v_{inf} = 1300$~km/s, $\Mdot = 3 \times 10^{-6} \Msol$/yr. The best fit gives the solution $a = 0.25$~au $=2.7 \Rsol$, $e = 0.20$, $i = 62 \adeg$ and $A = +146 \adeg$ (Fig.~\ref{fig:igrj16320}, central and right panels).

Thus, we find that the model reproduces well the orbital evolution of a NS surrounded by absorbing matter, and modulated by the stellar wind density profile, as viewed by the observer. In particular, it reproduces:
i.) the orbital modulation of $\nh$ with the sharp peak at phase = 0.47 (Fig.~\ref{fig:igrj16320}, top left panel);
ii.) the smooth evolution of the hard X-ray lightcurve (Fig.~\ref{fig:igrj16320}, bottom left panel); and
iii.) the phase shift of their maxima, with the maximum $\nh$ at phase = 0.47 as seen by the observer, and the maximum {\it Swift}/BAT counts at phase = 0.53 at the time when the NS crosses the periastron.

\section{``The Bad'': IGR~J16465-4507}

IGR~J16465-4507 is an obscured sgHMXB hosting a B0.5-1~Ib star, with $\nh = 10^{22} \cmmoinsdeux$, that we call ``The Bad'' source, because it is located inbetween the sgHMXB domain and the BeHMXB domain in the Corbet diagram (Fig.~\ref{fig:corbet}), with $\Pspin = 228$\,s and $\Porb = 30$\,days.

\begin{figure}
  \begin{center}
    \includegraphics[width=4.2in,angle=90]{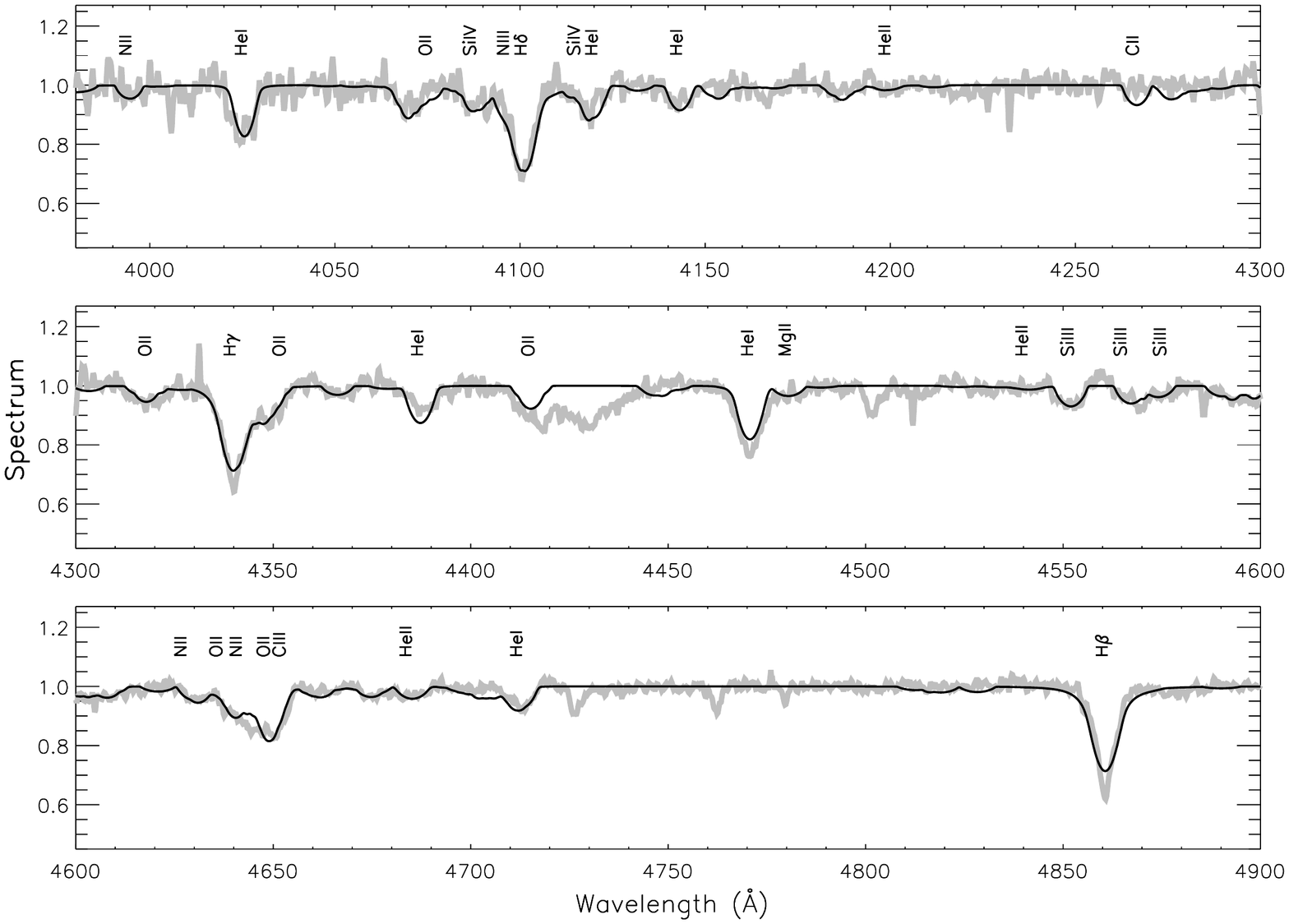}
    \caption{X-shooter spectrum of IGR J16465-4507 fitted with FASTWIND stellar spectral model. The spectrum is shown in grey, and the fit obtained is reported in black.}
    \label{fig:igrj16465}
  \end{center}
\end{figure}

We analysed all optical and infrared (OIR) photometric and spectroscopic observations obtained at ESO with EMMI, SUSI2, SOFI, FORS1 and X-shooter instruments, from 2006 to 2012. This analysis, reported in detail in \cite[Chaty et al. (2016)]{chaty:2016}, first allowed us to accurately establish the spectral type of the companion star to B0.5-1~Ib. Observations with X-shooter allowed us to get large-band spectra (UVB, VIS and NIR, reported in Fig.~\ref{fig:igrj16465}), to build a whole photometric and spectroscopic spectral energy distribution (SED), and to detect an IR excess, likely due to circumstellar cold material.

We then fitted absorption and emission lines using the stellar spectral model FASTWIND —-Fast Analysis of STellar atmospheres with WINDs--, a spherical, non-LTE model atmosphere code with mass loss and line-blanketing \cite[(Santolaya-Rey et al. 1997; Puls et al. 2005; Castro et al. 2012)]{santolaya:1997,puls:2005,castro:2012}. The result is reported in Fig.~\ref{fig:igrj16465}. We point out that it is one of the few sgHMXB for which the spectral type of the companion star has been accurately fitted with realistic stellar atmospheric spectral model.

From this fit, we derive that the supergiant star is characterized by a high rotation velocity: $v \times sin(i) = 320 \pm 8$~km/s, implying that this star must have a small radius (smaller than $15 \Rsol$), in order not to break up. We determine that $T_{eff} = 26 000$~K, with an envelope expanding at $v \sim 170$~km/s.

Looking at the Corbet diagram (Fig.~\ref{fig:corbet}), we see that IGR~J16465-4507 is located inbetween sgHMXB and BeHMXB domains, like two other supergiant fast X-ray transients (SFXT):
    IGR~J18483-0311 ($\Pspin = 21$\,s and $\Porb = 18.5$\,days)
    and IGR~J11215-5952 ($\Pspin = 187$\,s and $\Porb = 165$\,days),
    see \cite[Liu et al. (2011)]{liu:2011} and references therein.
    The question that arises is then: which process does make these SFXT look like BeHMXB?

To answer to this question, there are 2 possibilities:

i.) The high rotation velocity of these SFXT originated from the rapid rotator nature of O-type emission line stars (main sequence Oe-X-ray binary). In this case, these systems exhibit 2 periods of accretion during their evolution, first during the main sequence, and then during the supergiant phase \cite[(the NS would spin at its equilibrium period for a sgHMXB hosting a B1~Ia star, with $B = 3 \times 10^{12}$~G, Liu et al. 2011)]{liu:2011}.

ii.) Most SFXT tend to have an eccentric orbit, which would reduce the equilibrium period of the NS at the settling accretion stage, by a factor between 10 and 100 \cite[(Postnov et al. 2018)]{postnov:2018}, and explain the position of some sgHMXB in the BeHMXB domain. This is also visible in the eccentricity vs orbital period plot of BeHMXB, sgHMXB and SFXT reported in the extensive review of HMXB seen by INTEGRAL \cite[(Sidoli \& Paizis 2018)]{sidoli:2018}.

\section{``The Ugly'': IGR~J16318-4848}

IGR~J16318-4848 is a highly obscured sgHMXB hosting a sgB[e] star, with $\nh = 6 \times 10^{24} \cmmoinsdeux$, that we call ``The Ugly'' source, because with an average spin period typical of supergiant systems, it would be located inside the BeHMXB domain in the Corbet diagram (Fig.~\ref{fig:corbet}), with $\Porb = 80$\,days. Its companion star is a luminous sgB[e] star with a stratified circumstellar enveloppe, characterized by the following parameters: $L = 10^6 \Lsol$, $M = 30 \Msol$, $T = 22 000$~K, $R = 20 \Rsol = 0.1$~au.

We studied this source thanks to a full set of photometric and spectroscopic observations, obtained at ESO NTT/SOFI, VLT/VISIR, {\it Spitzer}, and {\it Herschel}. It exhibits strong emission lines, a MIR excess, an ionised wind at a velocity of 400 km/s, shocked [FeII] lines, some dense regions of circumstellar matter ($> 10^{5-6}$~cm$^{-3}$), and other lense dense regions, with NaI, Ne, Si, characteristic of a stratified envelope \cite[(Filliatre \& Chaty 2004, Chaty \& Rahoui 2012)]{filliatre:2004,chaty:2012a}.

\begin{figure}
  \begin{center}
    \includegraphics[width=5.25in,angle=0]{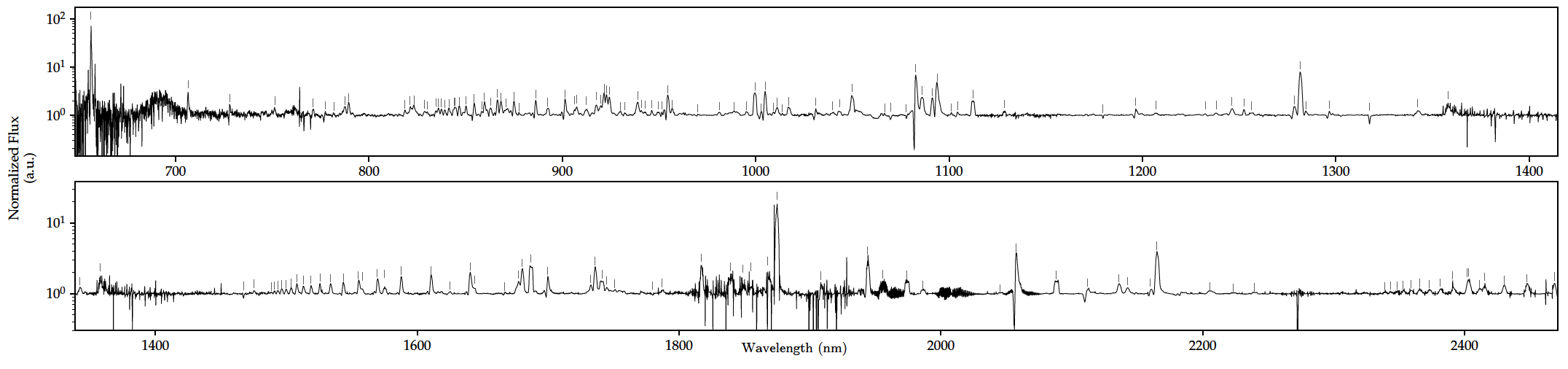}
    \includegraphics[width=5.25in,angle=0]{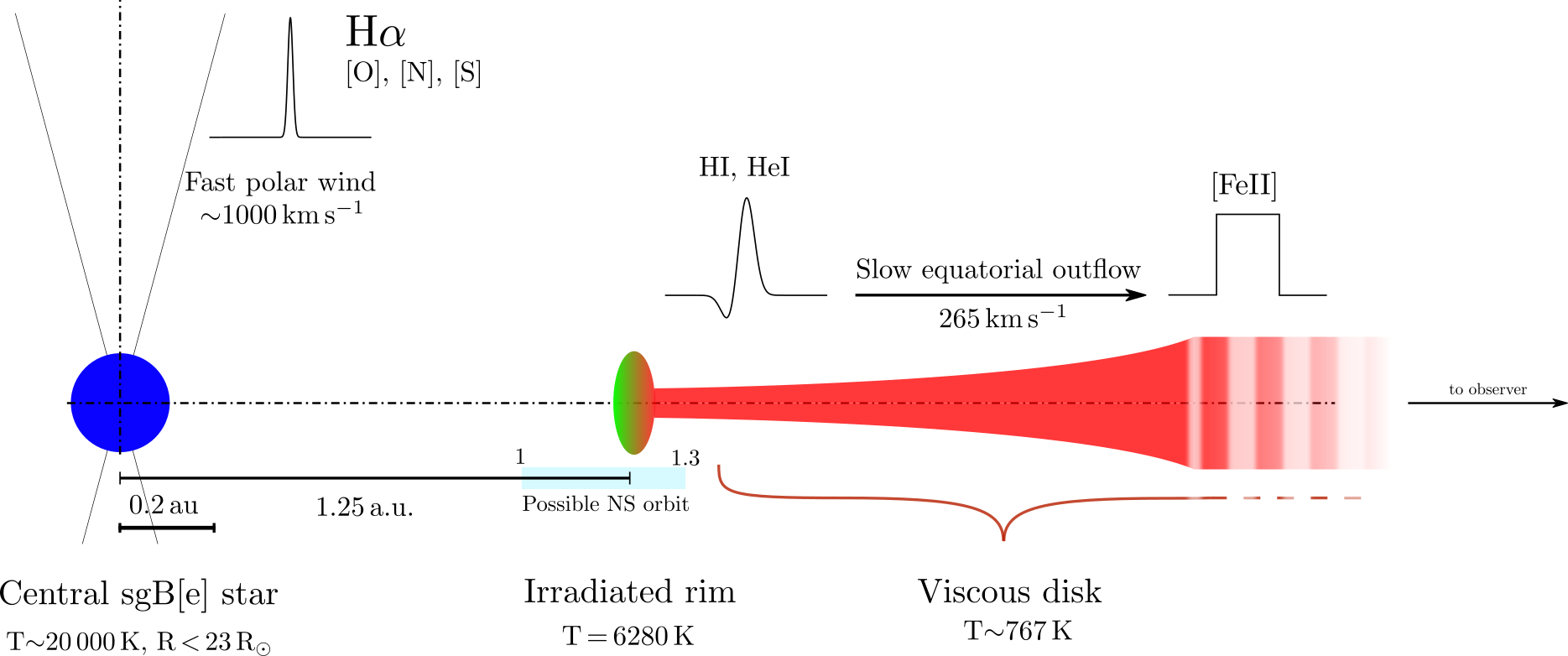}
    \caption{Top: ESO VLT/X-shooter UVB-VIS-NIR spectrum of IGR~J16318-4848. Bottom: Artist view of the scenario explaining the rare properties of IGR~J16318-4848.}
    \label{fig:igrj16318}
  \end{center}
\end{figure}

We obtained in 2012 an ESO VLT/X-shooter UVB-VIS-NIR spectroscopy, shown in Fig.~\ref{fig:igrj16318} (top). Preliminary results of the study of this spectrum are described in this volume \cite[(Fortin et al. 2018)]{fortin:2018c}. This spectrum shows prominent HI Balmer, Paschen, Bracket, and Pfund series, with a mean P-Cygni profile at 264 km/s. The Balmer H$\alpha$ line exhibits some evidence for a double component, which, when subtracting a mean P-Cygni profile obtained by stacking all HI lines, shows a narrow Balmer H$\alpha$ coming from a diluted and fast polar wind, suggesting that we see the binary system as edge-on. He lines also display P-Cygni profile at 164 km/s, with an extra broad emission at 840 km/s. Both H and He lines seem to come from the medium inbetween the supergiant star and the internal part of circumstellar material. The spectrum also shows forbidden [NII], [OI] and [SII] lines, likely coming from external part of the circumstellar material, characterized by a low density. In addition, we detect flat-topped profile of [Fe] lines, at a velocity of 262 km/s. These flat-topped forbidden lines are characteristic of lines formed inside a spherical shell of outflow and/or expanding material of low density: they are probably coming from a slow equatorial outflow, relatively far away from the star (see an artistic view of the scenario in Fig.~\ref{fig:igrj16318} bottom).

The scenario that we built to explain the rare properties of this system is based on Herbig Ae/Be model, with a torus geometry, an irradiated rim and a viscous disk extending up to $12 \Rstar$ \cite[(Chaty \& Rahoui 2012)]{chaty:2012a}. By taking the orbital period of $\Porb =80$~days (\cite[Iyer \& Paul 2017]{iyer:2017}), we can derive that the compact object orbits within the dense disk rim.

The binary system is currently transiting to Roche-lobe overflow, with a deep spiral-in, entering the Common Envelope Phase. Since the nature of the compact object is unknown, we have to consider two possibilities: i.) In case it is a NS, this short $\Porb$ HMXB will not survive the Common Envelope Phase, and becomes an ideal Thorne-Zytkow candidate (\cite[Tauris et al. 2017]{tauris:2017}). ii.) In case the compact object is a BH with a mass ratio $q<3.5$, the HMXB should survive the spiral-in phase, and then become a Wolf-Rayet X-ray binary with an orbital period of a few days, eventually leading to the merging of both compact objects \cite[(van den Heuvel et al. 2017)]{vandenheuvel:2017a}.

\section{Conclusions}

Concerning ``The Good'' source (IGR~J16320-4751), we performed a {\it Swift}-BAT and {\it XMM}-Newton study of accretion and absorption along the NS orbiting the supergiant star, and we were able to reconstruct the nature and parameters of the system, accurately fitting both $\nh$ and {\it Swift}-BAT counts.

For ``The Bad'' source (IGR~J16465-4507): our X-shooter spectrum, fitted on a stellar spectrum modeling allowed us to show that it was a rapidly rotating supergiant star, either descending from Oe (Be-like) rapid rotator, or an SFXT with a NS on an eccentric orbit.

We finally reported on a study of ``The Ugly'' source (IGR~J16318-4848), that we observed with various instruments, and among them the large-band ESO/VLT/X-shooter spectrograph. Our multi-spectrum analysis showed that IGR~J16318-4848 is an HMXB hosting a sgB[e] star, with a complex environment, including a stratified envelope, and a rim+disk surroundings.

\section*{Acknowledgments}

This work, supported by the Labex UnivEarthS programme of Université Sorbonne Paris Cité (Interface project I10 --From binaries to gravitational waves--, and by the Centre National d'Etudes Spatiales (CNES), was based on observations obtained with MINE --Multi-wavelength {\it INTEGRAL} NEtwork--.



\begin{discussion}

\discuss{Ileyk El Mellah}{These results are very interesting! Are you able to detect the accretion wake in the observations of the Good source?}

\discuss{Sylvain Chaty}{No, we can only derive the overall variation along the full orbit, but do not have the S/N high enough to get some additional details from the {\it Swift}-BAT lightcurve.}

\end{discussion}

\end{document}